\documentclass[aps,prb,twocolumn,amsmath,amssymb]{revtex4}
\usepackage{epsfig}
\usepackage{graphicx}
\usepackage{color}
\definecolor{Red}{rgb}{1,0,0}

\newcommand{\be}{\begin{equation}}

\newcommand{\ee}{\end{equation}}
\newcommand{\mean}[1]{\langle #1 \rangle}

\begin{document}

\title{Does a ferromagnet with spin-dependent masses produce a spin-filtering
effect in a ferromagnetic/insulator/superconductor junction?}
\author{Gaetano Annunziata, Mario Cuoco, Paola Gentile, Alfonso Romano, Canio Noce}
\affiliation{SPIN-CNR, I-84084 Fisciano (Salerno), Italy \\
Dipartimento di Fisica ``E. R. Caianiello'', Universit\`a di
Salerno, I-84084 Fisciano (Salerno), Italy}

\begin{abstract}
We analyze charge transport through a ballistic
ferromagnet/insulator/superconductor junction by means of the
Bogoliubov–-de Gennes equations. We take into account the
possibility that ferromagnetism in the first electrode may be
driven by a mass renormalization of oppositely polarized carriers,
i.e. by a spin bandwidth asymmetry, rather than by a rigid
splitting of up-and down-spin electron bands as in a standard
Stoner ferromagnet. By evaluating the averaged charge conductance
for both an $s$- and a $d_{x^2-y^2}$-wave order parameter for the
S side, we show that the mass mismatch in the ferromagnetic
electrode may mimic a spin active barrier. Indeed, in the $s$-wave
case we show that under suitable conditions the spin dependent
conductance of minority carriers below the energy gap $\Delta_0$
can be larger than for majority carriers, and lower above $\Delta_0$.
On the other hand, for a $d_{x^2-y^2}$-wave superconductor similar
spin-dependent effects give rise to an asymmetric peak splitting
in the conductance. These results suggest that the junction
may work as a spin-filtering device.
\end{abstract}

\date{\today}

\maketitle

\section{INTRODUCTION}

In recent years, the investigation of transport properties in
hybrid structures realized with a ferromagnet and a superconductor
separated by a thin insulating layer (F/I/S) has received
considerable theoretical and experimental attention. Indeed, the
study of these heterostructures may allow to probe relevant
properties of the materials constituting the junction, such as the
polarization of the ferromagnet~\cite{Soulen98,Mazin99} and/or the
gap amplitude and symmetry of the
superconductor.~\cite{Tedrow94,Bode03} Moreover, using these
junctions, potential device applications may be investigated
ranging from electronics to spintronics as well as to information
and communication technologies.~\cite{spintronicsreview} On the
other hand, the basic process occurring at the boundaries
separating the superconducting material and the normal metal, i.
e. the Andreev reflection~\cite{Andreev64} (AR), is strongly
sensitive to magnetic polarization. Furthermore, measurements of
the tunneling conductance in F/I/S  structures turn out to be an
important tool in understanding properties of old and new
superconducting materials, since this kind of measurement offers
the opportunity to probe the superconducting order parameter
symmetry in unconventional superconductors, the latter being ARs
phase sensitive. For instance, one of the strongest evidence
supporting $d$-wave symmetry for high-$T_c$ cuprates is the zero
bias conductance peak revealed in $ab$ plane tunneling conductance
from normal metal.~\cite{DeutRev}

To interpret the large amount of tunneling experiments performed
on F/I/S junctions, fundamental theories of transport, such as in
particular the one by Blonder, Tinkham and Klapwijk (BTK),~\cite{BTK}
have been suitably extended to take into account all possible
symmetries of the superconducting order
parameter.~\cite{TunnHF,TunnOrganics,TunnSrRuO,TunnFullerene,TunnBi,TunnFeAs}

Usually, the ferromagnetic side has predominantly been described
within the Stoner model, relying on the assumption that the bands
for different electron spin orientations have the same dispersion
and are rigidly shifted by the exchange interaction. Nevertheless,
the Stoner approach may be insufficient to describe some real
ferromagnets because many terms deriving from Coulomb repulsion
are eliminated from the full Hamiltonian.~\cite{Abinitio,Fazekas}
Moreover, some real systems such as the ferromagnetic transition
metals Fe, Co, and Ni and their alloys,~\cite{Wohlfarth80} weak
metallic ferromagnets such as ZrZn$_2$~\cite{ZrZnJap,ZrZnNature}
and Sc$_3$In,~\cite{Matthias61,HirschScIn} colossal
magnetoresistance manganites such as
La$_{1-x}$Sr$_x$MnO$_3$,~\cite{Schiffer95} and rare earth
hexaborides such as EuB$_6$~\cite{Matthias68,HirschEuB} require a
theoretical description going beyond the Stoner method. Therefore,
when theoretically modelling F/I/S junctions, it may be
interesting to assume for the magnetism in the F electrode
microscopic scenarios other than the Stoner one. Special interest
has been devoted to a form of itinerant ferromagnetism driven by a
gain in kinetic energy deriving from a spin dependent bandwidth
renormalization, or, equivalently, by an effective mass splitting
between up- and down-spin
carriers.~\cite{Zener51,Hirsch,Campbell88,Okimoto95,Higashiguchi05,McCollam05}
Furthermore, we mention that the interplay of superconductivity and this
kinetically driven ferromagnetism has been shown to originate
different features compared to the Stoner case, concerning the
phenomena of coexistence, proximity and
transport.~\cite{Cuoco,Annunziata}
Among the results obtained, we have shown that a F/I/S junction turns out to be a
valuable device to probe ferromagnetism, also providing
a way to estimate the possible mass mismatch of the
carriers. We have also investigated the properties of the same
junction, considering for the superconducting order
parameter a chiral $p$-wave and a $d$-wave symmetry. We have
found out that, as soon as a finite magnetization appears in the
F-side, very peculiar effects take place as a consequence of the
propagation in the ferromagnetic layer of superconducting
oscillating components of different symmetry compared to the bulk
one developing in the S-side. Moreover, the density of states at
the interface, in the $p$- as well as in the $d$-wave case,
exhibits distinct effects on the opening of gap-like structures in
the density of states of both kinds of spin species.

Motivated by the above considerations, we analyze here
the charge transport through a F/I/S junction where a spin
bandwidth asymmetry mechanism can be active in the ferromagnetic
layer. The motivation behind this choice is here related to the
investigation of the following issue: {\it can one observe in this
kind of junction a spin-dependent transport allowing to give rise
to spin-filtering effects?} We show that under specific conditions
this behavior is possible, as a consequence of the fact that the
mass mismatch makes up- and down-spin electrons see a different
height of the insulating barrier. The specific features of this
effect depend on the pairing symmetry of the superconducting order
parameter. In particular, here we consider the cases of a
conventional $s$-wave superconductor and of an unconventional
$d_{x^2-y^2}$-wave one.

The problem is here investigated by solving
the Bogoliubov–-de Gennes (BdG) equations~\cite{BdG} within an
extended BTK approach, formulated for a
two-dimensional F/I/US ballistic junction. We note that this method has been
generalized in the last years to take into account higher
dimensionalities, anisotropic forms of the superconducting order
parameter, different Fermi energies for the two sides of the
junction, and a spin-flip interfacial
scattering.~\cite{Kashiwaya99,ZuticValls,Dong,Zhu00,Zhu99,Barsic,Stefan1,Linder,deJong,
Tanaka_transp_d,Tanaka_transp_p,YokoyamaNCSC}

The paper is organized as follows. In Section II we introduce the
model and the related method of solution, while in Section III the
results correspondingly obtained are shown. Finally, the last section
is devoted to the conclusions.

\section{FORMULATION}

\begin{figure}
\begin{center}
\includegraphics[width=0.4\textwidth]{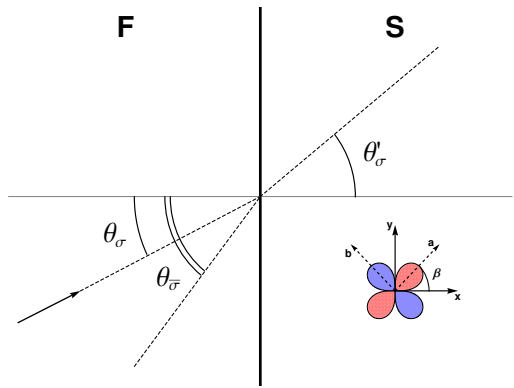}
\caption{(Color online) Schematic illustration of F/I/S junction
described in the text. Here, $\theta_\sigma$,
$\theta_{\bar{\sigma}}$, and $\theta_\sigma '$ are injection
Andreev reflection, and transmission angles for electrons and
quasiparticles with spin $\sigma$. $\beta$ is the angle formed by
the crystallographic $a$ axis of a $d$-wave superconductor with
the $x$ axis.} \label{sketch}
\end{center}
\end{figure}

Let us consider a two-dimensional F/I/S junction in the clean
limit, having the geometry shown in Fig.~\ref{sketch}. The
system is built up of two semi-infinite layers connected by an
infinitely thin insulating barrier described by a potential of the
form $V(\mathbf{r})=H\delta(x)$. The dimensionless parameter
$Z = 2 m' H /(\hbar^2 k'_F)$ everywhere in this paper will
conveniently characterize the strength of the interfacial
scattering. Here $m'$ and $k'_F$ are the quasiparticles
effective mass and the Fermi wave-vector for the superconductor,
respectively. The interface lies along the $y$ direction at $x=0$ so that the region $x<0$, the F side, is
occupied by an itinerant ferromagnet, while the region $x>0$, the
S side, is occupied by a singlet superconductor. The F side
may correspond to a standard Stoner ferromagnet (STF), to a spin
bandwidth asymmetry ferromagnet (SBAF), or to a system where both
kinds of microscopic mechanism are present.

The excitations propagating through the junction are obtained by
means of the single-particle Hamiltonian
\begin{eqnarray}
H_0^\sigma & = & \left[
-\hbar^2\mathbf{\nabla}^2/2m_\sigma-\rho_\sigma U
-E_F\right]\Theta(-x) \nonumber \\ && +
 \left[ -\hbar^2 \mathbf{\nabla}^2/2m'-E_F' \right]\Theta(x)+V(\mathbf{r})\; ,
\label{spHam}
\end{eqnarray}
where $\sigma=\uparrow,\downarrow$, $m_\sigma$ is the effective
mass for $\sigma$-polarized electrons in the F side,
$\rho_{\uparrow(\downarrow)}=+1(-1)$, $U$ is the exchange
interaction, $E_F$ is the Fermi energy of the ferromagnet,
$\Theta(x)$ is the Heaviside step function, and $E_F'$ is the Fermi energy for the
superconductor, respectively.  The BdG equations are
\be
\left( \begin{matrix} H_0^\sigma && \Delta \\
\Delta^\ast && -H_0^{\bar{\sigma}} \end{matrix} \right) \left(
\begin{matrix} u_\sigma \\ v_{\bar{\sigma}} \end{matrix} \right)
=\varepsilon \left( \begin{matrix} u_\sigma \\ v_{\bar{\sigma}}
\end{matrix} \right), \,\sigma=\uparrow,\downarrow\quad,
\label{eigensystem} \ee where $\bar{\sigma}=-\sigma$ and
$\left(u_\sigma,v_{\bar{\sigma}}\right)\equiv\Psi_\sigma$ is the
energy eigenstate in the electron-hole space associated with the
eigenvalue $\varepsilon$. Eqs.~(\ref{eigensystem}) admit an
analytical solution in the approximation of a rigid
superconducting pair potential, i.e.
$\Delta(\mathbf{r})=\Delta(\theta')\Theta(x)$, where $\theta'$ is
the angular variable for the S side (see Fig.~\ref{sketch}).
Neglecting proximity effect at the interface is a standard procedure
in the BTK approach which is known to give qualitative and quantitative
agreement with fully self-consistent calculation in the tunneling limit $Z\gg 1$.
On the other hand,  for high interface transparency some quantitative deviations can be found,
as shown for instance in Ref.~\onlinecite{Barsic} where the tunneling conductance
of clean ferromagnet/superconductor junctions has been  evaluated via
a fully self-consistent numerical solution of the microscopic BdG equations.
Since the Hamiltonian is invariant under translations along the
$y$-axis, we can rewrite the eigenstates of the model as
$\Psi_\sigma(\mathbf{r})=
e^{i\mathbf{k}_\parallel\cdot\mathbf{r}}\psi_\sigma(x)$, thus
reducing the effective dimensionality of the problem.

Various scattering processes are possible at the interface for an
electron injected from the F side with spin $\sigma$ and momentum
$\mathbf{k}_\sigma^+$
($k_\sigma^+=\left[\left(2m_\sigma/\hbar^2\right)\left(E_F+\rho_\sigma
U+\varepsilon\right)\right]^{1/2}$):

$a)$ normal reflection;

$b)$ AR resulting in a hole with momentum
$\mathbf{k}_{\bar{\sigma}}^-$ ($k_{\bar{\sigma}}^-=
\left[\left(2m_{\bar{\sigma}}/\hbar^2\right)\left(E_F+\rho_{\bar{\sigma}}
U-\varepsilon\right)\right]^{1/2}$) belonging to the opposite spin
band and a Cooper pair transmitted in the superconductor;

$c)$ transmission as an electron-like quasiparticle with momentum
$\mathbf{k}_\sigma '^+$ ($k_\sigma '^+= \left[\left(2m'/\hbar^2\right)\left(E_F'+\sqrt{\varepsilon^2-|\Delta_{\sigma +}|^2}\right)\right]^{1/2}$);

$d)$ transmission as a hole-like quasiparticle with momentum
$\mathbf{k}_\sigma '^-$ ($k_\sigma '^-=
\left[\left(2m'/\hbar^2\right)\left(E_F'-\sqrt{\varepsilon^2-|\Delta_{\sigma -}|^2}\right)\right]^{1/2}$).

\noindent Here, $\Delta_{\sigma \pm}=|\Delta_{\sigma \pm}| \ e^{i \phi^{\pm}_{\sigma}}$ is the pair potential felt by electron-like ($+$) and hole-like ($-$) quasiparticles. We note that the spin dependence of $\Delta_{\sigma \pm}$ comes out from
the different trajectories followed by up- and down-spin quasiparticles.
The processes actually taking place depend on the energy, momentum
and spin orientation of the incoming electrons, as well as on the
interfacial barrier strength, the polarization in the F side and
the symmetry of the superconducting order parameter in the S side.

For standard low-biased F/I/S junctions one has
$E_F,E_F'\gg(\varepsilon,|\Delta|)$, so that one can apply the
Andreev approximation~\cite{Andreev64} and fix the momenta on the
Fermi surfaces. Thus, the solutions of BdG equations can be
written as
\begin{eqnarray}
\psi_\sigma^F(x)=e^{ik_{\sigma,x}^Fx}\left(\begin{matrix} 1\\0
\end{matrix} \right)+ a_\sigma
e^{ik_{{\bar{\sigma},x}}^Fx}\left(\begin{matrix} 0\\1 \end{matrix}
\right)+
b_\sigma e^{-ik_{\sigma,x}^Fx}\left(\begin{matrix} 1\\0 \end{matrix} \right)\\
\psi_\sigma^S(x)=c_\sigma e^{ik_{\sigma,x}'^Fx}\left(
\begin{matrix} u_+\\
e^{-i\phi_+}v_+
\end{matrix} \right)+ d_\sigma
e^{-ik_{\sigma,x}'^Fx}\left(\begin{matrix}
e^{i\phi_-}u_-\\
v_- \end{matrix} \right)
\end{eqnarray}
where
$$
u_\pm  = \sqrt{\frac{\varepsilon\pm\sqrt{\varepsilon^2-|\Delta_{\sigma \pm}|^2}}{2\varepsilon}}
$$
$$
v_\pm  = \sqrt{\frac{\varepsilon\mp\sqrt{\varepsilon^2-|\Delta_{\sigma \pm}|^2}}{2\varepsilon}}
\; .
$$
The superscript F in the wavevectors denotes that they are taken on
the Fermi surfaces.

The probability amplitude coefficients
$a_\sigma$, $b_\sigma$, $c_\sigma$, $d_\sigma$  for the scattering
processes may be calculated from the boundary conditions
\begin{eqnarray}
\psi_\sigma^F(0)=\psi_\sigma^S(0) \nonumber \\ \left.\frac{d
u_\sigma^F}{d x}\right|_{x=0}- \left.\frac{m_\sigma}{m'}\frac{d
u_\sigma^S}{d
x}\right|_{x=0}&=&-\frac{2 H\ m_\sigma}{\hbar^2}u_\sigma^S(0) \nonumber \\
\left.\frac{d v_{\bar{\sigma}}^F}{d x}\right|_{x=0}-
\left.\frac{m_{\bar{\sigma}}}{m'}\frac{d v_{\bar{\sigma}}^S}{d
x}\right|_{x=0}&=&-\frac{2 H\
m_{\bar{\sigma}}}{\hbar^2}v_{\bar{\sigma}}^S(0) \; .
\label{raccordo}
\end{eqnarray}

\noindent We point out that the mass asymmetry between up- and down-spin
electrons enters explicitly the above equations. This is at the
origin of the different behavior across the barrier of the two
kinds of carrier, that will be discussed in detail in the next
Section.

The charge differential conductance at $T=0$ is calculated from
the ratio between the charge flux across the junction and the
incident flux at that bias, and may be easily evaluated from the
probabilities associated with the four processes listed
above.
We proceed exactly as in Ref.~\onlinecite{ZuticValls}, firstly calculating the charge
differential conductance for each spin channel
\be
G_\sigma(\varepsilon,\theta)= P_\sigma \left(
1+\frac{k_{\bar{\sigma},x}^F}{k_{\sigma,x}^F} \
|a_\sigma(\varepsilon,\theta)|^2-|b_\sigma(\varepsilon,\theta)|^2
\right) \, ,
\ee
where $\theta$ is the angle formed by the
momentum of the electrons propagating from the F side with respect
to the normal to the interface (see Fig.~\ref{sketch}), and
$P_\sigma=n_\sigma/(n_\uparrow+n_\downarrow)$ is the fraction of
electrons occupying the $\sigma$-spin band of the metallic
ferromagnet.
Then, using the averaged differential conductance for given spin orientation
\be
\mean{G_\sigma(\varepsilon)}=\int_{-\theta_{C}^\sigma}^{\theta_{C}^\sigma}
d\theta \ \cos \theta \ G_\sigma(\varepsilon,\theta) /
\int_{-\theta_{C}^\sigma}^{\theta_{C}^\sigma} d\theta \ \cos
\theta  ,
\ee
where $\theta_{C}^\sigma$ is the critical angle for the transmission of {$\sigma$-spin} electrons,
we get the averaged charge conductance as
\be
\mean{G(\varepsilon)}  =  \mean{G_\uparrow(\varepsilon)}+
\mean{G_\downarrow(\varepsilon)}\, . \label{charge cond}\\
\ee

\section{RESULTS}
As already mentioned in the previous Sections, we consider a F/I/S
junction where ferromagnetism in the F-side may also be originated
by a relative change in the bandwidths of electrons with opposite
spin. In particular, we investigate the effect on the transport
properties of the junction of the spin-dependent barrier
renormalization stemming from the boundary conditions in Eqs.(5),
also making a comparison with the conventional case of a Stoner
ferromagnet. As far as the latter is concerned, it has been
shown,~\cite{Kashiwaya99} in the case of a $d$-wave
superconducting electrode, that when ferromagnetism is described
in terms of the Stoner model and a spin active barrier is
considered, the conductance spectra for majority (minority) spins
are shifted downwards (upwards) in energy. Besides, the spin
polarization in the ferromagnet induces an imbalance of the peak
magnitude and position, which can be considered as a signal of the
spin polarization.

\begin{figure}[!th]
\begin{center}
\includegraphics[width=0.5\textwidth]{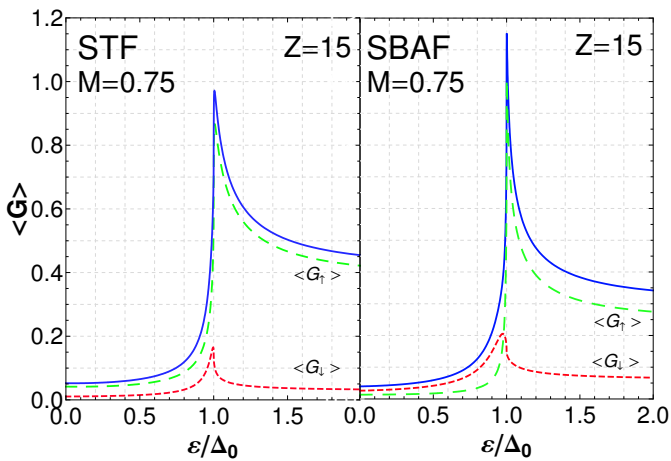}
\caption{(Color online) Spin-dependent charge conductances and
total averaged conductance (full lines) for a F/I/S junction
plotted as functions of the positive bias $\varepsilon/\Delta_0$.
The left and the right figures refer to the cases of a Stoner-like
ferromagnet ($U/E_F=0.75, m_\uparrow / m_\downarrow=1$) and of a
spin band asymmetry ferromagnet ($U/E_F=0, m_\uparrow /
m_\downarrow=7$), respectively. In both cases we have fixed $Z=15$
and $M=0.75$.}
\end{center}
\end{figure}

\begin{figure}[!th]
\begin{center}
\includegraphics[width=0.45\textwidth]{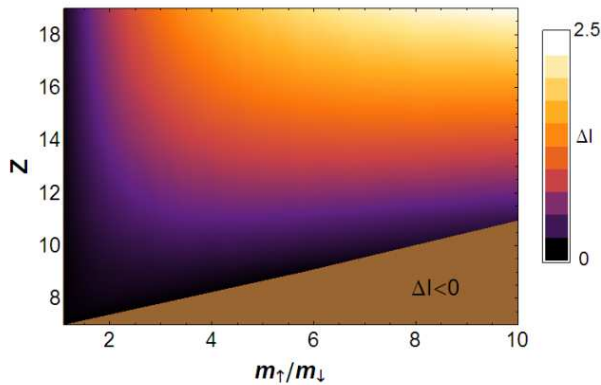}
\caption{(Color online) Density plot showing the quantity $\Delta
I$, defined in Eq. (9), as a function of the barrier height $Z$
and the mass ratio $m_\uparrow / m_\downarrow$. Positive values of
$\Delta I$ correspond to values of the minority spin integrated
conductance, i. e. up to $\Delta_0$, larger than the corresponding
majority ones. Negative values of $\Delta I$  indicate that minority down-spin
electrons contribute to the current less than majority up-spin
ones. Colors are associated with the values taken by the quantity
$\Delta I$. }
\end{center}
\end{figure}

We start by considering the case of an isotropic $s$-wave
superconductor (i. e. $\Delta_{\sigma \pm}=\Delta_0$) coupled to a
STF or a SBAF, assuming a moderately large value of the interface
barrier strength $Z$ (=15) and a fixed value of the
magnetization $M$ (=0.75) (we refer here to the same definition of
$M$ introduced in Ref.~\onlinecite{Annunziata}). From the behavior
of the averaged charge conductance, reported in Fig.~2 together
with its two spin-resolved contributions, we see that in the SBAF
case (right panel) for energies lower than the ground-state gap
value $\Delta_0$, the component $<G_{\downarrow}>$ associated with
the minority down-spin carriers is larger than the component
$<G_{\uparrow}>$ associated with the majority up-spin ones, this
order relation being reversed for $\varepsilon$ larger than
$\Delta_0$, approximately. This behavior strongly contrasts with
the one observed for the STF case (left panel), where
$<G_{\downarrow}>$ is always lower than $<G_{\uparrow}>$. A
possible explanation of this apparently anomalous behavior may be
that in the SBAF case the barrier acquires a spin active character
and thus the up-spin electrons, although larger in number than the
down-spin ones, feel a stronger value of the barrier height,
resulting in a correspondingly reduced charge conductance. We
point out that the manifestation of this behavior depends on the
actual values of $Z$ and the mass ratio
$m_{\uparrow}/m_{\downarrow}$. To clarify this point, in Fig.~3 we
report a density plot for the quantity
\be
\Delta I = \frac{<I_{\downarrow}> - < I_{\uparrow}>}{<
I_{\uparrow}>}
\ee
\noindent as a function of $Z$ and $m_{\uparrow}/m_{\downarrow}$,
where we have defined the spin-dependent integrated conductance
\[
<I_\sigma>=\int_0 ^{\Delta_0} <G_{\sigma}(\varepsilon)> d\varepsilon\, .
\]
From this figure we deduce that above moderately high values
of the bare barrier height, a low degree of the mass mismatch is
already sufficient to give rise to a charge current contribution
from minority spin larger than the one produced by
majority spin ones.

\begin{figure}[!th]
\begin{center}
\includegraphics[width=0.45\textwidth]{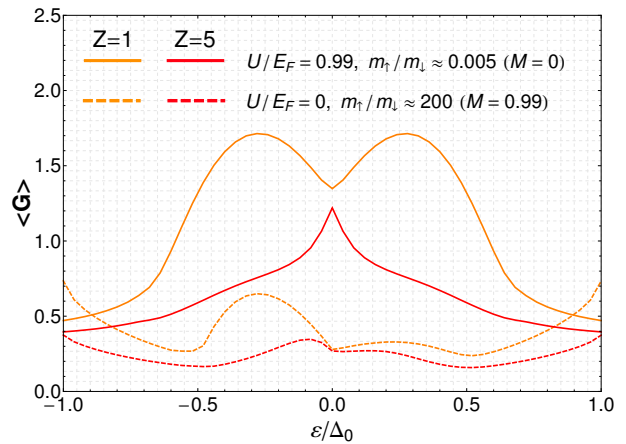}
\caption{(Color online) Averaged charge
conductance curves plotted as a function of the renormalized
energy ($\varepsilon/\Delta_0$) for two values of the barrier
height $Z$, namely $Z$=1 (orange curves) and $Z$=5 (red curves).
The full lines refer to a N/I/S junction, where the magnetization
$M$=0, whereas the dotted ones correspond to a F/I/S junction with
a strong mass mismatch effect and a very large magnetization
$M$=0.99. In both cases we consider a $d_{x^2-y^2}$-symmetry superconductor.}
\end{center}
\end{figure}

\begin{figure}[!th]
\begin{center}
\includegraphics[width=0.45\textwidth]{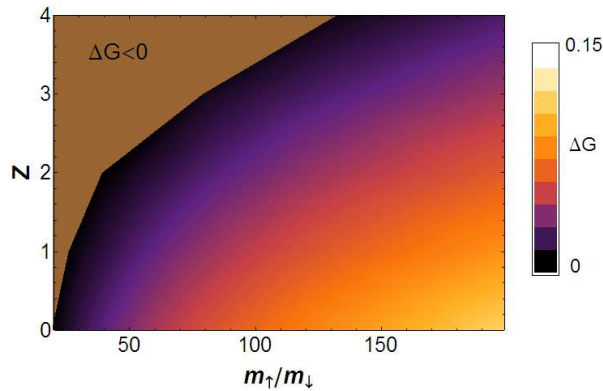}
\caption{(Color online) Density plot showing the
values of $Z$ and $m_\uparrow / m_\downarrow$ for which in a
SBAF/I/S junction an asymmetric splitting of zero bias peak
appear as a signature of spin--active interface. Colors are
associated with the values taken by the discrete right derivative
at the origin $\Delta G=(<G(\varepsilon = 0.1\Delta_0)>-<G(0)>)/<G(0)>$.}
\end{center}
\end{figure}

The results obtained in the case of a $d_{x^2-y^2}$-wave
superconductor (i. e. $\Delta_{\sigma,\pm}=\Delta_0 \cos
[2(\theta_\sigma'\mp\beta)]$ with $\beta=\pi/4$) are
presented in Fig.~4. The full lines correspond to the case where
the F-side is normal, i. e. the magnetization in the ferromagnet
is zero as a consequence of the combined counteracting effect of
the two microscopic mechanisms, realized here with a suitable
combination of the exchange interaction and a vanishingly small
value of the mass spin asymmetry ratio. From this figure we
observe a symmetric behavior with respect to zero bias of the
averaged charge conductance, with the peak splitting disappearing
as the barrier height is increased. In terms of the spin resolved
averaged conductances, this behavior is easily explained observing
that $<G_{\uparrow}>$ and $<G_{\downarrow}>$ coincide for not too
small $Z$, becoming symmetrically shifted with respect to each
other as $Z$ is decreased, as a consequence of the energy gain for
majority spin, and the energy loss for minority ones, produced by
the spin sensitive tunneling processes.

The case of an F/I/S junction with a SBAF is analyzed here
assuming a value of $m_{\uparrow}/m_{\downarrow}$ large enough to
make the ferromagnet behave almost like an half-metal. This choice
is motivated by the fact that we want to emphasize the role played
by the mass mismatch on the charge conductance. The asymmetry in the
charge conductance profile that we observe in the SBAF case (dotted lines in Fig.~4)
follows the behavior of the spin-resolved charge conductance contributions $<G_{\uparrow}>$ and $<G_{\downarrow}>$, not shown here for brevity, characterized by peaks of different heights, which moreover are not symmetrically shifted with respect to zero energy. For $M\neq 0$ the relation
$<G_{\uparrow}(\varepsilon)>=<G_{\downarrow}(-\varepsilon)>$ is thus no
more valid, so that the symmetric behavior observed before for
$M=0$ is lost and the total averaged charge conductance
$<G>=<G_{\uparrow}>+<G_{\downarrow}>$ exhibits an asymmetric
shape.

These results show remarkable similarities to the ones reported in
Ref.~\onlinecite{Kashiwaya99}, and thus support our conjecture
that a mass mismatch ferromagnet produces effects which are
qualitatively similar to the ones obtained with spin active
barriers. Of course, as previously stated, in our case no
spin-flip process takes place in the insulating layer, with the
transport becoming spin dependent only because up- and down-spin
electrons see a different effective value of the barrier height
$Z$, as a consequence of their mass mismatch. We also stress that
the large value of $m_{\uparrow}/m_{\downarrow}$ $(\sim 10^2$)
required to get a spin active behavior in the two-dimensional case
analyzed here, gets considerably reduced in three dimensions.
Actually, we have verified that for the same case of large
magnetization considered in Fig.~4, it becomes lower by about one
order of magnitude ($m_{\uparrow}/m_{\downarrow}\simeq 30$).
We remark that also in the case of a $d_{x^2-y^2}$-wave superconductor,
the results depend on the choice of the microscopic parameters of the model.
To clarify this
point in Fig.~5 we draw a contour plot giving the values assumed
by the discrete right derivative of the averaged charge
conductance at the origin, namely $\Delta G=(<G(\varepsilon =
0.1\Delta_0)>-<G(0)>)/<G(0)>$, as a function of $Z$ and
$m_{\uparrow}/m_{\downarrow}$. We infer that in the limit of low
transparency of the barrier ($Z\geq$ 3.5), $\Delta G$ is always
negative indicating that the peak in the averaged conductance is
located at zero bias, independently of the value assumed by the
mass ratio. On the other hand, for high barrier transparency, the
subtle interplay between $Z$ and $m_{\uparrow}/m_{\downarrow}$
gives rise to the peak splitting discussed above, which is here
signalled by a positive value of $\Delta G$.

As a final consideration, we note that the features observed
when a $d_{x^2-y^2}$-wave superconductor is considered are
visible only in the limit of high transparency because in the
tunneling limit only zero-energy Andreev bound states contribute
to transport and this is unaffected by magnetization and/or spin
active scattering at the interface, as pointed out in
Ref.~\onlinecite{Kashiwaya99}.

\section{CONCLUSIONS}

We have investigated the effects induced by a ferromagnet with
spin-dependent masses in a F/I/S junction, evaluating the averaged
charge conductance for both an $s$- and a $d_{x^2-y^2}$-wave
superconducting order parameter for the S side. We have shown that
the mass mismatch in the ferromagnetic electrode leads to new
features in the transport properties, allowing to conclude that
the presence of this kind of ferromagnet may mimic the behavior
of a spin active barrier. This is achieved as a consequence of the fact that the
mass asymmetry between up- and down-spin electrons entering
explicitly the boundary condition equations, make electrons with
opposite spin feel different values of the barrier height,
eventually resulting in a spin-filtering effect. When the S-side
is an isotropic $s$-wave superconductor, we have shown that for
biases lower than the ground-state gap amplitude $\Delta_0$, the
conductance associated with the minority-spin carriers is larger
than for the majority-spin ones, this ordering being reversed when
biases larger than $\Delta_0$ are considered. This result suggests
that a junction with a mass mismatch ferromagnet induces an
effective spin-active interfacial effect. Nevertheless, this
result requires a cooperative effect between the barrier and
the mass mismatch: above moderately high barrier height, low
values of the mass ratio are already able to produce a
minority-spin charge conductance component higher than the
corresponding majority-spin one. When a $d_{x^2-y^2}$-wave order
parameter for the S side is considered, an asymmetry in the
averaged charge conductance is found, whereas if the F layer
becomes normal a symmetric behavior of this quantity comes out.
Also in this case, the asymmetry we find depends on the
interplay between the barrier height and the mass mismatch, and it
may disappear in specific regions of the corresponding parameter
space. We also emphasize that the asymmetric peak splitting in
the charge conductance in the case of $d_{x^2-y^2}$-wave S side or
the knowledge of $\Delta I$ for a standard $s$-wave superconductor
may potentially be used to design spin-filtering
devices, with the degree of mass mismatch being tuned by suitable
choices of the SBAF.

We notice that the tunneling conductance of clean  ferromagnet/superconductor junctions may be evaluated also via a fully  self-consistent numerical solution of the microscopic Bogoliubov-de Gennes equations.~\cite{Barsic} Within this method, it is found that the features of the  conductance curves look similar to those obtained via non-self-consistent procedures. Nevertheless, some differences emerge when the dependence of the charge conductance on the Fermi wave-vector mismatch and exchange field parameters is considered. For instance, in the non-self-consistent approach $G(V)$ is analytically found to be independent of the Fermi wave-vector mismatch at the gap edge, while this quantity shows a monotonically increasing trend when the self-consistent procedure is applied.  Moreover, differently from the non-self-consistent results the subgap conductance is reduced for smaller Fermi wave-vector mismatch and larger exchange field.  These findings indicate that, while the non-self-consistent approach is a good tool to understand qualitatively some of the relevant features of the transport in a ferromagnet/superconductor junction, a fully self-consistent approach is needed to properly model experimental data.

Additionally, we point out that though in our calculation there is no Fermi wave-vector mismatch, an analogous role is played by the spin-mass mismatch assumed to be responsible for the ferromagnetism of the F side of the junction. Apart from possibly being at the origin of the microscopic mechanism of spin polarization, a difference in the masses of up- and down-spin electrons gives rise to peculiar effects, discussed extensively in this paper, producing a "true" spin-filtering barrier. In this way, an effective scattering potential near the interface, similar to the one generated by the Fermi wave-vector mismatch, is created, despite the fact that our calculations are performed in a non-self-consistent way. The value of the mass ratio, together with the one of the exchange energy, allows in this way an analysis of the experimental data certainly more reliable than standard one-parameter models do.

We are confident that in the tunneling limit ($Z\gg1$)  our non-self-consistent results may look similar to those obtained within a self-consistent approach, whereas in the high transparency limit the situation will be different. Nonetheless, previous results obtained in this limit within a microscopic self-consistent approach~\cite{Cuoco08} suggest that the proximity effect is affected by the ferromagnet assumed in the F side, giving rise to different outcomes when a STF or a SBAF is considered,  these effects being again not related to the Fermi wave-vector mismatch.
To better clarify these aspects, a study of the transport phenomena in a SBAF/superconductor junction within a self-consistent approach is in progress.

Finally, we mention that it has been shown that the spin-filtering
properties of the Eu chalcogenides, used together with
one-electron quantum dots, have been proposed as the basis for a
method to convert single-spin into single-charge measurements, by
means of single-electron transistors or quantum point
contacts.~\cite{DiVinc,spintronicsreview} Thus, an F/I/S junction with a SBAF
may represent an important tool for the manipulation of the spin
degrees of freedom in solid-state systems, as concerns both
spin transport and spin relaxation phenomena.
We notice that this kind of junction may be realized using as F electrode
ferromagnetic manganites or semiconducting ferromagnets with impurity
bands, with the S side being made by an high-T$_c$ superconductor.
We believe that the findings reported in this paper
can turn out to be relevant for the experimental probe
of specific features of magnetic and superconducting materials.
Indeed, the polarization of itinerant or semiconductor ferromagnets can be estimated
by fitting the experimental data from point contact or scanning tunneling
microscopy measurements within the BTK approach, treating the exchange splitting and mass mismatch
as independent fitting parameters.
They can also prove to be useful in spintronic
applications and devices requiring an efficient way of controlling
separately charge and spin currents. Moreover, the
knowledge of the charge response at different values of the spin
polarization may be used to perform high-sensitive magnetization
measurements.  Given the increasing number of experimental investigations in this
rapidly growing field, we believe that our results may
provide a useful contribution to the comprehension of
some relevant phenomena involving spin polarized tunneling
spectroscopy

\section*{ACKNOWLEDGMENTS}

We wish to thank sincerely Prof. Jacob Linder for useful discussions.


\begin{thebibliography}{99}
\bibitem{Soulen98} R. J. Soulen Jr., J. M. Byers, M. S. Osofsky, B. Nadgorny,
T. Ambrose, S. F. Cheng, P. R. Broussard, C. T. Tanaka, J. Nowak,
J. S. Moodera,    A. Barry, and J. M. D. Coey, Science {\bf 282},
85 (1998); S. K. Upadhyay, A. Palanisami, R. N. Louie, and R. A.
Buhrman, Phys. Rev. Lett. {\bf 81}, 3247 (1998).
\bibitem{Mazin99} I.I. Mazin, Phys. Rev. Lett. {\bf 83}, 1427
(1999).
\bibitem{Tedrow94} R. Meservey and P.M. Tedrow, Phys. Rep. {\bf
238}, 173 (1994).
\bibitem{Bode03} M. Bode, Rep. Prog. Phys. {\bf 66}, 523 (2003).
\bibitem{spintronicsreview} I. \v{Z}uti\'c, J. Fabian, and S.D. Sarma,
Rev. Mod. Phys. {\bf 76}, 323 (2004).
\bibitem{Andreev64} A. F. Andreev, Zh. Eksp. Teor. Fiz. {\bf 46}, 1823 (1964) [Sov. Phys. JETP {\bf 19}, 1228 (1964)].
\bibitem{DeutRev} G. Deutscher, Rev. Mod. Phys. {\bf 77}, 109 (2005).
\bibitem{BTK} G. E. Blonder, M. Tinkham, and T. M. Klapwijk, Phys. Rev. B {\bf 25}, 4515 (1982).
\bibitem{TunnHF} I. Iguchi, T. Yasuda, Y. Nuki, and T. Komatsubara, Phys. Rev. B {\bf 35},
8873 (1987).
\bibitem{TunnOrganics} C. More, G. Roger, J.P.
Sorbier, D. Jérome, M. Ribault, and K. Bechgaard, J. Physique
Lett. {\bf 42}, 313 (1981); M. Yoshimura, H. Shigekawa, H. Nejoh,
G. Saito, Y. Saito, and A. Kawazu, Phys. Rev. B {\bf 43}, 13590
(1991).
\bibitem{TunnSrRuO} M. D. Upward, L. P. Kouwenhoven, A. F.
Morpurgo, N. Kikugawa, Z. Q. Mao, and Y. Maeno, Phys. Rev. B {\bf
65}, 220512(R) (2002).
\bibitem{TunnFullerene} Z. Zhang, C.-C. Chen, S.P.
Kelty, H. Dai, and C.M. Lieber, Nature {\bf 353}, 333 (1991).
\bibitem{TunnBi} Q. Huang, J. F. Zasadzinski, N.
Tralshawala, K. E. Gray, D. G. Hinks, J. L. Peng, and R. L.
Greene, Nature {\bf 347}, 369 (1990).
\bibitem{TunnFeAs} T. Y. Chen, Z. Tesanovic, R. H. Liu, X. H.
Chen, and C. L. Chien, Nature (London) {\bf 453}, 1224 (2008).
\bibitem{Abinitio} R. Strack, and D. Vollhardt, Phys. Rev. Lett. {\bf 72}, 3425
(1994); M. Kollar, R. Strack, and D. Vollhardt, Phys. Rev. B {\bf
53}, 9225 (1996); D. Vollhardt, N. Bl\"umer, K. Held, J. Schlipf,
and M. Ulmke, Z. Phys. B {\bf 103}, 283 (1997); J. Wahle, N.
Bl\"umer, J. Schlipf, K. Held, and D. Vollhardt, Phys. Rev. B {\bf
58}, 12749 (1998).
\bibitem{Fazekas} P. Fazekas, in {\it Lecture Notes on Electron
Correlation and Magnetism} (North-Holland, Amsterdam, 1980).
\bibitem{Wohlfarth80} E. P. Wohlfarth, in {\it Ferromagnetic Materials},
edited by E. P. Wohlfarth (North-Holland, Amsterdam, 1980).
\bibitem{ZrZnJap} S. Ogawa and N. Sakamoto, J. Phys. Soc. Japan {\bf 22}, 1214 (1967).
\bibitem{ZrZnNature} C. Pfleiderer, M. Uhlatz, S. M. Hayden, R. Vollmer, H. v. L\"ohneysen,
N. R. Berhoeft, and G. G. Lonzarich, Nature(London) {\bf 412}, 58
(2001).
\bibitem{Matthias61} B. T. Matthias, A. M. Clogston, H. J. Williams,
E. Corenzwit, and R. C. Sherwood , Phys. Rev. Lett. {\bf 7}, 7
(1961); B. T. Matthias and R. M. Bozorth, Phys. Rev. {\bf 109},
604 (1958); S. G. Mishra, Mod. Phys. Lett. B {\bf 4}, 83 (1990).
\bibitem{HirschScIn} J.E. Hirsch, Phys. Rev. B {\bf44}, 675 (1991).
\bibitem{Schiffer95} P. Schiffer, A. P. Ramirez, W. Bao,
and S.-W. Cheong, Phys. Rev. Lett. {\bf 75}, 3336 (1995).
\bibitem{Matthias68} B.T. Matthias, T.H. Geballe, K. Andres,
E. Corenzwit, G.W. Hull, and J.P. Maita, Science {\bf 159}, 530
(1968); Z. Fisk, D. C. Johnston, B. Cornut, S. von Molnar, S.
Oseroff, and R. Calvo, J. Appl. Phys. {\bf 50}, 1911 (1979); L.
Degiorgi, E. Felder, H. R. Ott, J. L. Sarrao, and Z. Fisk, Phys.
Rev. Lett. {\bf 79}, 5134 (1997).
\bibitem{HirschEuB} J.E. Hirsch, Phys. Rev. B {\bf 59}, 436 (1999).
\bibitem{Zener51} C. Zener, Phys. Rev. {\bf 82}, 403 (1951);
P.W. Anderson and H. Hasegawa, Phys. Rev. {\bf 100}, 675 (1955);
P.G. de Gennes, Phys. Rev. {\bf 118}, 141 (1960).
\bibitem{Hirsch} J.E. Hirsch, Phys. Rev. B {\bf 40}, 2354 (1989);
J.E. Hirsch, {\it ibid.} {\bf 40}, 9061 (1989); J.E. Hirsch, {\it
ibid.} {\bf 43}, 705 (1991); J.E. Hirsch, {\it ibid.} {\bf 59},
6256 (1999); J.E. Hirsch, {\it ibid.} {\bf 62}, 14131 (2000); J.E.
Hirsch, Physica C {\bf 341-348}, 211 (2000).
\bibitem{Campbell88} D. K. Campbell, J. T. Gammel, and E. Y. Loh, Phys. Rev. B {\bf 38}, 12043 (1988);
{\bf 42}, 475 (1990); S. Kivelson, W.-P. Su, J. R. Schrieffer, and
A. J. Heeger, Phys. Rev. Lett. {\bf 58}, 1899 (1987).
\bibitem{Okimoto95} Y. Okimoto, T. Katsufuji, T. Ishikawa,
A. Urushibara, T. Arima, and Y. Tokura, Phys. Rev. Lett. {\bf 75},
109 (1995); Y. Okimoto, T. Katsufuji, T. Ishikawa, T. Arima, and
Y. Tokura, Phys. Rev. B {\bf 55}, 4206 (1997); S. Broderick, B.
Ruzicka, L. Degiorgi, H. R. Ott, J. L. Sarrao, and Z. Fisk, Phys.
Rev. B {\bf 65}, 121102 (2002).
\bibitem{Higashiguchi05} M. Higashiguchi, K. Shimada, K. Nishiura, X. Cui, H. Namatame,
and Masaki Taniguchi, Phys. Rev. B {\bf 72}, 214438 (2005).
\bibitem{McCollam05} A. McCollam, S. R. Julian, P. M. C. Rourke, D. Aoki,
and J. Flouquet, Phys. Rev. Lett. {\bf 94}, 186401 (2005); I.
Sheikin, A. Gr\"oger, S. Raymond, D. Jaccard, D. Aoki, H. Harima,
and J. Flouquet, Phys. Rev. B {\bf 67}, 094420 (2003).
\bibitem{Cuoco} M. Cuoco, A. Romano, C. Noce,
and P. Gentile, Phys. Rev. B {\bf 78}, 054503 (2008); Z.-J. Ying,
M. Cuoco, C. Noce, and H.-Q. Zhou, Phys. Rev. B {\bf 78}, 104523
(2008); M. Cuoco, P. Gentile, and C. Noce, Phys. Rev. Lett. {\bf
91}, 197003 (2003).
\bibitem{Annunziata} G. Annunziata, M. Cuoco, C. Noce, A. Romano,
and P. Gentile, Phys. Rev. B {\bf 80}, 012503 (2009).
\bibitem{BdG} P. G. de Gennes, in {\it Superconductivity of
Metals and Alloys}, (W.A. Benjamin, Inc. New York, 1966).
\bibitem{Kashiwaya99} S. Kashiwaya, Y. Tanaka, N. Yoshida, and M. R. Beasley,
Phys. Rev. B {\bf 60}, 3572 (1999).
\bibitem{ZuticValls} I. \v{Z}uti\'c and O.T. Valls, Phys. Rev. B {\bf 60}, 6320 (1999);
I. \v{Z}uti\'c and O.T. Valls, {\it ibid.} {\bf 61}, 1555 (2000).
\bibitem{Dong} Z. C. Dong, D. Y. Xing, Z. D. Wang, Z. Zheng, and J. Dong,
Phys. Rev. B {\bf 63}, 144520 (2001).
\bibitem{Zhu00} J.-X. Zhu and C.S. Ting, Phys. Rev. B {\bf 61}, 1456 (2000).
\bibitem{Zhu99} J.-X. Zhu, B. Friedman, and C. S. Ting, Phys. Rev. B {\bf 59}, 9558
(1999).
\bibitem{Barsic} P. H. Barsic and O. T. Valls, Phys. Rev. B {\bf 79}, 014502 (2009).
\bibitem{Stefan1} N. Stefanakis and R. M\'elin, J. Phys.: Condens. Matter {\bf 15}, 3401 (2003);
N. Stefanakis and R. M\'elin, {\it ibid.} {\bf 15},
4239 (2003).
\bibitem{Linder} J. Linder and A. Sudb\o, Phys. Rev. B {\bf 75}, 134509
(2007); J. Linder and A. Sudb\o, {\it ibid.} {\bf 79} 020501
(2009).
\bibitem{deJong} M. J. M. de Jong and C. W. J. Beenakker, Phys. Rev. Lett. {\bf 74}, 1657 (1995).
\bibitem{Tanaka_transp_d} Y. Tanaka, Yu. V. Nazarov, and S. Kashiwaya, Phys. Rev. Lett. {\bf 90}, 167003 (2003);
Y. Tanaka, Yu. V. Nazarov, A. A. Golubov, and S. Kashiwaya, Phys.
Rev. B {\bf 69}, 144519 (2004).
\bibitem{Tanaka_transp_p} Y. Tanaka and S. Kashiwaya, Phys. Rev. B {\bf 70}, 012507 (2004);
Y. Tanaka, S. Kashiwaya, and T. Yokoyama, Phys. Rev. B {\bf 71}, 094513
(2005).
\bibitem{YokoyamaNCSC} T. Yokoyama, Y. Tanaka, and J. Inoue, Phys. Rev. B {\bf 72}, 220504
(2005).
\bibitem{Cuoco08} M. Cuoco, A. Romano, C. Noce, and P. Gentile, Phys. Rev. B {\bf 78}, 054503 (2008).
\bibitem{DiVinc} D. P. Di Vincenzo,  J. Appl. Phys.{\bf 85}, 4785 (1999).

\end{thebibliography}
\end{document}